\documentclass[prl,twocolumn,showpacs,preprintnumbers,amsmath,amssymb]{revtex4}
\usepackage{graphicx}

\usepackage{dcolumn}
\usepackage{bm}

\begin{document}

\title{Could the Casimir Effect explain the Energetics of
High-Temperature Superconductors?}

\author{Achim Kempf}
\affiliation{
Departments of Applied Mathematics and Physics, University of Waterloo\\
and Perimeter Institute for Theoretical Physics\\ Waterloo, Ontario N2L
3G1, Canada}


\begin{abstract}
It is proposed that the Casimir energy of the electromagnetic field in
between the layers of cuprate superconductors could significantly
contribute to the energy balance in these materials. Since the Casimir
energy is strongly dependent on the distance between the layers a
corresponding dependence is predicted for the superconducting transition
temperatures. Comparison with the experimental data on the transition
temperatures of epitaxial superlattices of alternating layers of YBCO and
PrBCO of varying thickness shows that these are well reproduced.
\end{abstract}

\pacs{74.72.-h, 74.20.-z,12.20.-m}

\maketitle The question has remained open how Cooper pairs can be stable
at around 100K where some high temperature superconductors (HTSCs) are
still superconducting. In particular, the phonon-mediated attractive
electron-electron interaction of BCS theory is known to be too weak at
these temperatures, see, e.g., \cite{micro,bcs}.

For a new approach to this problem, let us reconsider a feature that HTSCs
have in common, namely parallel superconducting layers which are separated
by layers of essentially insulating material. Since in between any two
conducting surfaces there occurs a Casimir effect, see \cite{casimir}, the
effect also occurs between the parallel superconducting layers in HTSCs,
as was first pointed out in \cite{ak}. Before estimating its significance
in HTSCs, let us briefly review the Casimir effect's underlying mechanism.

We begin by recalling that the ground state energy of a quantum harmonic
oscillator is not necessarily fixed. It can be lowered, for example, by
coupling the oscillator to degrees of freedom which decrease the amplitude
of its zero-point fluctuations. This happens, e.g., when these extra
degrees of freedom effectively increase the oscillator's mass. An example
would be the vibrational oscillator of a diatomic molecule that captures a
neutron. Similarly, the ground state energy of the electromagnetic field
can be lowered by suppressing the zero point fluctuations of some of its
electromagnetic field modes. As Casimir showed, this happens, for example,
when the electromagnetic field couples to conducting charge carriers which
are confined to two parallel plates, in which case certain modes in the
$c$-direction become suppressed. In the case of HTSCs, as the temperature
is lowered below $T_c$, superconducting charge carriers form in parallel
layers and they therefore lower the electromagnetic ground state energy.
Our aim here is to estimate if this lowering of the energy might be
sufficient to make the formation of Cooper pairs energetically favorable
at temperatures as high as 100K.


The derivation of the Casimir effect for two conducting objects from first
principles requires the calculation of the ground state energy of the
quantum system that consists of both, the electromagnetic field and the
charge carriers. This calculation is hard and it is instructive to
consider first the simplified case where the two objects are what may be
called \it ideal \rm conductors, namely conductors whose conductivity and
Mei{\ss}ner effect expel electromagnetic fields of all wavelengths with
vanishing penetration depths. In this case, the charge degrees of freedom
are effectively integrated out and one is left with the noninteracting
quantum field theory of electromagnetism together with the boundary
condition that the electromagnetic fields vanish at the objects' surfaces.
Only those electromagnetic modes which obey this boundary condition
contribute their ground state energy. In this way, the total ground state
energy of the system depends on the objects' (shape and) distance, $a$.
The distance dependence of the energy implies a force, the Casimir force.
These calculations require renormalization but are straightforward for
simple geometries such as two parallel plates. The Casimir force between
two ideally conducting, neutral and parallel plates of large area $A$ and
distance $a$ is given by:
\begin{equation}
F(a) = - \frac{\pi^2\hbar cA}{240a^4} \label{casimir}
\end{equation}
Corrections that take into account the finite conductivity of real metals
have been calculated for geometries such as parallel plates and a plate
and a sphere, along with corrections for finite surface roughness and
finite temperature, see \cite{casimir-reviews}. Recent experiments
measured the Casimir force between a metallic plate and sphere down to
distances around $100nm$, confirming the theoretical predictions to a
precision of $0.5\%$ \cite{mohideen-exp}.

From Eq.\ref{casimir}, the ground state energy of the quantum system
consisting of the electromagnetic field and two ideally conducting,
neutral and parallel plates reads:
\begin{equation} E(a) = -
\frac{\pi^2\hbar cA}{720a^3} \label{casimir2}
\end{equation}
The integration constant is chosen such that the energy of the system
vanishes at infinite plate separation, or equivalently, in the absence of
conducting plates. We see that a system of two conducting parallel plates
at a distance $a$ is energetically lower than the same system of two
parallel plates at distance $a$ if the two plates are insulators. Indeed,
if two parallel insulating plates possess a microscopic mechanism which
allows them to create ideally conducting charge carriers at an energy
expense of no more than $E(a)$ (and correspondingly less at finite
temperature), then the two plates are energetically driven to use this
microscopic mechanism to create ideally conducting charge carriers.

This scenario may apply to HTSCs. They possess parallel layers which are
initially insulating but can become superconducting at low enough
temperature, i.e., there exists some microscopic mechanism that allows
these layers to create superconducting charge carriers. The formation of
the Cooper pairs leads not to ideal conductivity but to superconductivity,
which in turn leads to a partial suppression of the fluctuations of
certain electromagnetic modes. This leads, therefore, to some lowering of
the vacuum energy of the electromagnetic field. If this lowering of the
electromagnetic zero-point energy at the onset of superconductivity is
large enough then the Cu-O layers' initially non-superconducting charge
carriers are indeed energetically driven to utilize the available
microscopic mechanism in order to become Cooper pairs. We notice here the
importance of the non-superconducting state being an insulating state.
This is because the size of the change of the electromagnetic ground state
energy is crucial and it depends on the difference in conductivity in the
two phases.

In this scenario, it would be the very effects of superconductivity which
enable and stabilize superconductivity. Cooper pairs would derive their
stability collectively, across layers. Namely, Cooper pairs would be
stable because if sufficiently many of the Cooper pairs on opposing layers
were to break up then the suppression of some electromagnetic modes'
zero-point fluctuations would cease and their electromagnetic ground state
energy would have to go back up to its unsuppressed level.

While a Casimir effect must occur in HTSCs it is, \it  a priori, \rm not
clear if it is indeed strong enough to lead to this scenario, especially
because the superconducting Cu-O layers are less efficient at suppressing
electromagnetic modes than ideal conductors would be. Let us, therefore,
consider the well-studied material $YBa_2Cu_3O_{7-x}$ (YBCO) which becomes
superconducting at around $92K$. The crystallographic unit cell contains
two copper oxide layers at a distance of $a_b \approx 0.39nm$ and
neighboring such bi-layers are separated by a layer of essentially
nonconducting material of width $a_i \approx 1.17nm$. The area density of
superconducting charge carriers on each individual Cu-O layer may reach on
the order of $n_s \approx 10^{14}/cm^2$.

For our purposes, the case of YBCO is of particular interest because of
the availability of experimental data, \cite{too,budai}, on epitaxial
superlattices in which slabs of YBCO alternate with slabs of insulating
material, namely $PrBa_2Cu_3O_{7-x}$ (PrBCO). For example, in the
experiments reported in \cite{budai}, the authors varied the thickness of
the YBCO slabs from $M=1$ to $M=8$ unit cells and the thickness, $a_m$, of
the PrBCO slabs from $N=1$ to $N=16$ unit cells, i.e., in the range
$a_m=2$nm to $20$nm. The superconducting transition temperature was
measured as a function, $T_c(N,a_m)$, of $N$ and $a_m$. The results are
summarized in Fig.3 of \cite{budai}. Since $T_c$ is known to be a function
of the layer separation and since the Casimir effect is sensitive to
varying layer separations those data provide a good testing ground for the
present ansatz.

Let us initially consider the simplified case in which the Cu-O layers in
their superconducting state are taken to be ideal conductors in the above
sense, which means that they cause electromagnetic fields of all
wavelengths to vanish on their surface. We also assume that the Cu-O
layers are of negligible thickness and separated by vacuum. In this case,
the charge degrees of freedom are effectively integrated out and we are
left with the free quantum field theory of electromagnetic fields with the
boundary condition that these fields vanish on each of the superconducting
Cu-O layers. The ground state energy of the electromagnetic field between
any two ideally conducting layers of distance $a$ is then lowered by the
amount given in Eq.\ref{casimir2} and we can apply this result to all the
inter-layer distances that occur in the superlattice.

Concretely, each period of the superlattice contains a slab of $M$ unit
cells of YBCO (each cell with one bi-layer of Cu-O), followed by a slab of
insulating PrBCO of thickness $a_m$. Therefore, in each period of the
superlattice, the case that two neighboring superconducting Cu-O layers
are separated by the distance $a_b$ occurs $M$ times. The case that two
neighboring superconducting Cu-O layers are separated by the distance
$a_i$ occurs $(M-1)$ times and finally the case that two neighboring
superconducting Cu-O layers are separated by the distance $a_i+a_m$ occurs
once per period of the superlattice. The reduction of the ground state
energy of the combined electromagnetic and charge carrier system, within
the volume given by one period of the superlattice times an area $A$ in
the $ab$ plane, is therefore given by:
\begin{equation}
E^\text{(period)} = - \frac{ \pi^2\hbar
cA}{720}\left(\frac{M}{a_b^3}+\frac{M-1}{a_i^3} + \frac{1}{a_m^3}\right)
\end{equation}
Since one period of the superlattice contains $2 M$ layers of Cu-O, this
energy reduction is shared by $2 M n_s A $ superconducting charge
carriers, yielding the gap energy:
\begin{equation}
2\Delta = \frac{ E^\text{(period)}}{2MA~n_s}
\end{equation}
In HTSCs, the value of the variable $\eta$,
\begin{equation}
\eta =\frac{2\vert\Delta\vert}{k_BT_c},
\end{equation}
is thought to be around or somewhat larger than the BCS value of
$\eta\approx 3.5$. We obtain for the temperature $T_c$ of the
superconducting transition:
\begin{equation} T_c(M,a_m) = \frac{
\pi^2\hbar c}{1440~M n_s \eta ~k_B
}\left(\frac{M}{a_b^3}+\frac{M-1}{a_i^3} + \frac{1}{a_m^3}\right)
\label{ideal}
\end{equation}
\begin{figure}[h]
    \centerline{\includegraphics[width=230pt,angle=0]{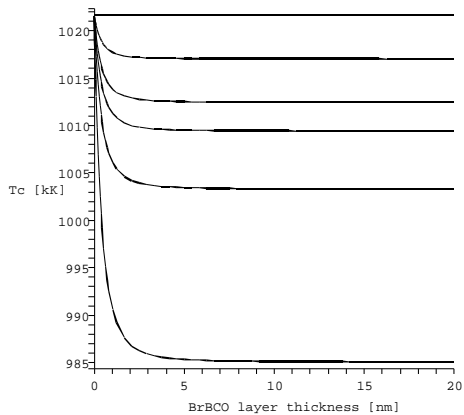}}
    \caption{\label{F:example0} $T_c$ for the ideal
    Casimir effect and $\eta=3.5,~n_s=5\times 10^{13}/cm^2$.
    The curves are,
    from bottom to top, for YBCO layer thicknesses
    $M=1,2,3,4,8,\infty$ (the latter is the
    case of pure YBCO). Notice the
    temperature scale of kilo Kelvin.}
\end{figure}
As Fig.\ref{F:example0} shows, the so-predicted transition temperature
curves are qualitatively of the right shape when compared with Fig.3 of
\cite{budai}. However, the transition temperatures are four orders of
magnitude too high! The onset of ideal conductivity on these Cu-O planes
would provide far more Casimir energy than is needed to explain the Cooper
pair binding energy. But the predicted Casimir effect for real Cu-O layers
is of course weaker.

The expression Eq.\ref{casimir2} for the Casimir energy function must be
corrected, for example, to take into account that the space between the
Cu-O layers is dielectric rather than a vacuum. It is also necessary to
correct for the non-flatness of the Cu-O layers and for finite temperature
effects. Most importantly, however, real superconducting Cu-O layers are
not ideal conductors in the sense above. Real Cu-O layers merely suppress
the penetration of electromagnetic fields and they do so less and less
efficiently the shorter the wavelength. In particular, modes whose
wavelengths in the $c$ direction are shorter than a certain length scale
contribute to the ground state energy essentially the same amount whether
the Cu-O planes are in their insulating or in their superconducting state.
These modes do not contribute to the Casimir effect, i.e., to the lowering
of the ground state energy at the onset of superconductivity in Cu-O
layers.

Thus, while the Casimir energy, $E(a)$, of the idealized case given in
Eq.\ref{casimir2} diverges as $E(a)\propto a^{-3}$ for $a\rightarrow 0$,
the corrected Casimir energy function $E_\text{corr}(a)$ must flatten at
some critical length, say $a_c$. For $a\rightarrow 0$, it is to be
expected that $E_\text{corr}(a)\rightarrow 0$ because the volume in which
the lowering of the density of the electromagnetic ground state energy
occurs, namely the volume between the plates, goes to zero as
$a\rightarrow 0$.

\begin{figure}[h]
    \centerline{\includegraphics[width=210pt,angle=0]{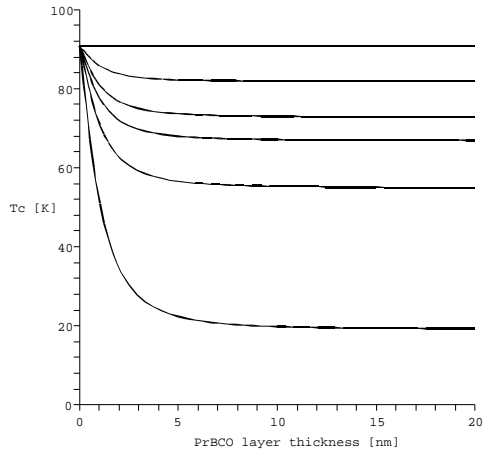}}
    \caption{\label{F:example1} $T_c$ predicted for
    corrected Casimir effect with the cutoff function $f_1(a) =e^{-a_c/a}$
    and setting $a_c=2.7nm$,
    $\eta=9.5,~n_s=9\times 10^{14}/cm^2$.
The curves are, from bottom to top, for YBCO slab thicknesses of
$M=1,2,3,4,8,\infty$.}
\end{figure}

\begin{figure}[h]
    \centerline{\includegraphics[width=210pt,angle=0]{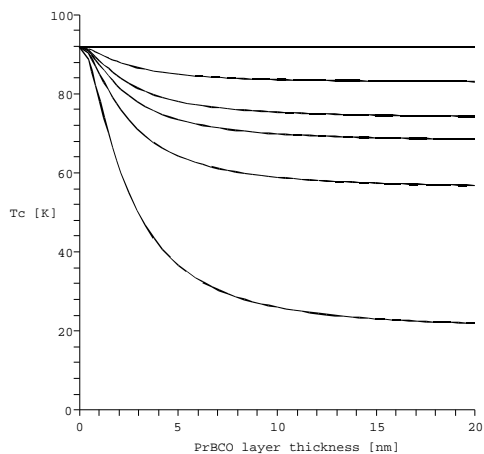}}
    \caption{\label{F:example2} $T_c$ predicted for
    corrected Casimir effect with the
    cutoff function $f_2(a) =e^{-\sqrt{a_c/a}}$ and setting $a_c=45nm$,
    $\eta=3.5,~n_s=5\times 10^{13}/cm^2$. The curves are, from bottom
     to top, for YBCO slab thicknesses of $M=1,2,3,4,8,\infty$.}
\end{figure}

The critical length, $a_c$, should be roughly in the range
$10^0...10^2nm$. Namely, while $a_c$ cannot be smaller than the coherence
length in the $c$ direction, $a_c$ also should not be much larger than
about $100nm$. The latter bound is suggested by the fact that measurements
of the Casimir effect between non-superconducting metal objects have shown
that the Casimir force persists even there, without changing sign, at
least down to values of $a\approx 100nm$, see \cite{mohideen-exp}. Let us
write the corrected expression $E_\text{corr}(a)$ as:
\begin{equation}
E_\text{corr}(a) = - \frac{\pi^2\hbar cA}{720~a^3}~f(a) \label{casimir3}
\end{equation}
Here, $f$ is a cutoff function of which we know that it obeys
$f(a)\rightarrow 0$ probably at least as fast as $a^3$ for $a\ll a_c$, and
we make the simplifying assumption that $f(a)\approx z$ for $a\gg a_c$
with $z=1$. It is important to calculate $f$ for concrete materials but
this is hard as it involves the renormalization of relativistic quantum
electrodynamics in a highly nontrivial background. For now, our aim is to
establish only roughly the potential role of the Casimir effect in HTSCs
and for this limited purpose the precise behavior of $f$ should not
matter. Also, while the value of $z$ is likely smaller than $1$, any value
of $z$ is straightforward to accommodate in our subsequent calculations.
Below, we will therefore consider the simple example $f_1(a) =
e^{-a_c/a}$, and for comparison also the example $f_2(a) =
e^{-\sqrt{a_c/a}}$ which describes a slightly softer cutoff. In a further
simplification, we treat the Casimir effects of neighboring pairs of
layers as independent so that we can then use $f$ to calculate
$E_\text{corr}^{(\text{period})}$ as before. We obtain for the critical
temperature:
\begin{equation}
T_c(M,a_m) = \frac{ \pi^2\hbar c\left(f(a_b)+(1-1/M)f(a_i) +
f(a_m)/m\right)}{1440 ~n_s \eta ~k_B}\label{x}
\end{equation}
At this point, both $\eta$ and $n_s$ are still free parameters, but we
notice that only their product enters in Eq.\ref{x}. The question is,
therefore, if, given a cutoff function, all six experimental curves
reported in Fig.3 of \cite{budai} can be reproduced at once when
appropriately choosing the two free parameters $a_c$ and $\eta\cdot n_s$.
That these six curves can in fact be reproduced surprisingly well is shown
in Figs.\ref{F:example1},\ref{F:example2}.
The comparatively large magnitude of the ideal Casimir effect (i.e. the
case $a_c=0$), see Fig.\ref{F:example0}, shows that, in principle, there
is in this scenario no energetic barrier to raising $T_c$ to significantly
higher temperatures. The challenge is to lower $a_c$, i.e., to find
layered materials, such as cuprates, for which the Casimir effect is as
large as possible. To this end, the material must normally be a very good
insulator while in the superconducting state the superconducting layers
must be able to efficiently suppress electromagnetic fields down to an as
small as possible wavelength, $a_c$.

Of course, $T_c$ could also be increased by reducing the layer separations
$a_b$ and $a_i$. For example, $T_c$ should generally increase with
pressure in the $c$-direction, except for situations where $a_b$ and $a_i$
become significantly smaller than $a_c$, in which case the absolute value
of the Casimir energy drops and therefore $T_c$ should drop as well. This
might occur, e.g., when the number of Cu-O layers per unit cell is
increased. A detailed comparison with data, see, e.g., \cite{mello-mult},
\cite{chakravarty} is in progress.

Depending on $f$ and the relative sizes of $a_b,a_i$ and $a_c$, the
crystal should also exhibit a finite expansion or contraction in the $c$
direction as it is cooled from just above to just below $T_c$, due to the
onset of the Casimir force between Cu-O layers. E.g., YBCO's elastic
modulus of $E \approx 10^{2}$GPa yields positive $\Delta L/L\approx
O(10^{-4})$ to $O(10^{-5})$ for the case of $f$ and $a_c$ as in Figs.2,3.
This effect could be measurable.

In addition, seemingly paradoxical situations are to be expected, where a
lower charge carrier density leads to a higher $T_c$, or to phenomena
related to a pseudo-gap. This is because when there are fewer charge
carriers, the Casimir energy can be shared among fewer of them, thereby
increasing the Cooper pair binding energy available to each pair. This is
counteracted by the fact that the Casimir energy only materializes as long
as the superconducting charge carrier density maintains a suppression of
the electromagnetic field in the Cu-O planes that is sufficient to yield a
Casimir effect.

We close with a remark on superconducting carbon nanotubes (CN), see
\cite{cn1}. From the above, it is plausible that, due to the Casimir
effect, $T_c$ may be higher for multi-walled than for single-walled CNs.
Indeed, very recently, a 30-fold increase in $T_c$ for multi-walled CNs
has been reported \cite{cn2}.

In conclusion, we have shown that the Casimir effect could play a
significant role in HTSCs, possibly even accounting for the overall
energetic stability of the Cooper pairs. The question as to the detailed
microscopic binding mechanism for Cooper pairs would still remain largely
open, however. We could only say that a) the stability of Cooper pairs
would be a collective phenomenon as it involves mode suppression across
layers and b) that since the Casimir effect concerns the energy stored in
electromagnetic modes, the relativistic retardation of the electromagnetic
interaction would play a role in HTSCs. Our calculations are simple enough
to be qualitatively robust. Quantitatively, however, the validity of our
results hinges on the assumptions made about the cutoff function $f$ and
on the assumption that the Casimir effects of neighboring pairs of layers
can be treated as independent, i.e., that these Casimir energies are
additive. These points will need to be investigated from first principles.
\medskip\newline \bf Acknowledgement: \rm This work has been supported by
CFI, OIT, PREA and the Canada Research Chairs Program of the National
Science and Engineering Research Council of Canada.

\vfill
\end{document}